\documentclass[12pt,letterpaper]{article}        
\pdfoutput=1
\usepackage[usenames,dvipsnames]{color}
\usepackage{jheppub,comment,bbm}

\DeclareRobustCommand{\spy}{\text{\reflectbox{$\beta$}}}

\title{\Large A holographic perspective on phonons and pseudo-phonons}

\author[a,c]{Andrea Amoretti,}
\author[b]{Daniel Are\'an,}
\author[c]{Riccardo Argurio,}
\author[d]{Daniele Musso}
\author[e]{and Leopoldo A. Pando Zayas}
\newcommand{\be}{\begin{equation}}
\newcommand{\ee}{\end{equation}}
\newcommand{\bea}{\begin{eqnarray}}
\newcommand{\eea}{\end{eqnarray}}

\newcommand{\nn} {\nonumber}

\affiliation[a]{Institute of Theoretical Physics and Astrophysics, 
University of W\"{u}rzburg, 97074 W\"{u}rzburg, Germany}
\affiliation[b]{Max-Planck-Institut f\"{u}r Physik (Werner-Heisenberg-Institut), F\"{o}hringer Ring 6, D-80805, Munich, Germany}
\affiliation[c]{Physique Th\'{e}orique et Math\'{e}matique and International Solvay Institutes
Universit\'{e} Libre de Bruxelles, C.P. 231, 1050 Brussels, Belgium}
\affiliation[d]{Departamento de F\'{i}sica de Part\'{i}culas, Universidade de Santiago de Compostela and Instituto
Galego de F\'{i}sica de Altas Enerx\'{i}as (IGFAE), E-15782, Santiago de Compostela, Spain.}
\affiliation[e]{Michigan Center for Theoretical Physics, Department of Physics, University of Michigan, Ann
Arbor, MI 48109, USA}
\emailAdd{Andrea.Amoretti@ulb.ac.be, darean@mpp.mpg.de, rargurio@ulb.ac.be, daniele.musso@usc.es, lpandoz@umich.edu} 

\abstract{We analyze the concomitant spontaneous breaking of translation and conformal symmetries
by introducing in a CFT a complex scalar operator that acquires a spatially dependent expectation value. 
The model, inspired by the holographic Q-lattice, provides a privileged setup to study the emergence
of phonons from a spontaneous translational symmetry breaking in a conformal field theory 
and offers valuable hints for the treatment of phonons in QFT at large.
We first analyze the Ward identity structure by means of standard QFT techniques, considering both 
spontaneous and explicit symmetry breaking.
Next, by implementing holographic renormalization, we show that the same set of Ward identities 
holds in the holographic Q-lattice. 
Eventually, relying on the holographic and QFT results,
we study the correlators realizing the symmetry breaking pattern and how they encode  
information about the low-energy spectrum.}

\preprint{MCTP-16-29, MPP-2016-336}

\begin{document}

\maketitle
\pagestyle{plain} \setcounter{page}{1}
\newcounter{bean}
\baselineskip16pt

\section{Introduction and motivation}

Despite being the Goldstone modes associated with broken spatial translations,
phonons are rarely described as emerging genuinely from a symmetry breaking process.
In general the effective field theories containing phonons assume the presence of a lattice
without accounting for its spontaneous formation. The description of the renormalization group flow
of a microscopic translational invariant theory that develops a lattice in the infrared
is complicated, for instance, because it entails many-body physics. The characteristics of the symmetry 
breaking do affect, however, the construction of the low-energy effective field theory 
(see for instance \cite{Nicolis:2015sra,Nicolis:2013lma}).
The simplest example being that a proper counting of the Goldstone modes is crucial
to build the correct effective field theory; the case of phonons 
has already been addressed, for example, in \cite{Leutwyler:1993gf}.

Ward identities are the cornerstone of the UV/IR connection implied by a symmetry breaking process. 
They contain information about the symmetries of the microscopic theory 
and, at the same time, about how the dynamically generated scales enter the low-energy correlators.
Note that the Ward identities contain only kinematic information as opposed to dynamical.
They describe the interplay of the various scales in the system without quantitatively predicting 
their value. As an intuitively clear example,
the Ward identity analysis of the spontaneous 
breaking of a global U$(1)$ symmetry predicts the possibility of superfluidity without 
implying the actual existence of superfluids. The latter requires the explicit solution 
of a model dynamically realizing the symmetry breaking pattern. Nevertheless, the Ward identities
state \emph{a priori} that the would-be superfluid is described at low energy by the massless 
Goldstone mode associated to the broken U$(1)$ (see \emph{e.g.}~\cite{Son:2002zn}).

In this work we put forward a novel approach to the realization of phonons as Goldstone modes
based on the analysis of the Ward identities. In particular, we focus on the concomitant
breaking of translational and conformal symmetry
due to a single complex scalar operator acquiring a space-dependent vacuum expectation 
value. The modulation of the scalar operator is characterized by a wave-vector $k^\mu$ and realizes a lattice 
along one spatial direction. 
This is the minimalist scenario implementing the pattern of symmetry breaking 
we are interested in. It allows us to show explicitly the emergence of the dilaton and phonon modes
and discuss their modifications upon adding an explicit component to the breaking (\emph{i.e.} a nontrivial source).

We perform our analysis in a generic quantum field theory (QFT) with the symmetry breaking 
pattern above, and complement
it with a concrete holographic realization consisting  in a Q-lattice model~\cite{Donos:2013eha}.
Holography constitutes a privileged setup that encodes the Ward identity structure of a QFT and 
allows the analysis of its renormalization group flow departing from a UV conformal fixed 
point~\cite{Bianchi:2001de,Bianchi:2001kw,Skenderis:2002wp,Papadimitriou:2010as,Papadimitriou:2011qb}. 
Moreover, our holographic analysis provides an interesting perspective on the problem at hand and yields useful input for our generic QFT setup.
Additionally, the holographic Q-lattice  belongs to the widely studied class of AdS/CMT models 
featuring translational 
symmetry breaking and clarifies some issues about the presence and characteristics of phonons
in this 
context~\cite{
  OoguriPark,Vegh:2013sk,Davison:2013txa,Blake:2013bqa,Blake:2013owa,Andrade:2013gsa,Donos:2014uba,Amoretti:2014zha,Donos:2014cya,Amoretti:2014mma,Baggioli:2014roa,Amoretti:2015gna,Alberte:2015isw}.

We also show that a general symmetry breaking pattern with both explicit and spontaneous 
components leads to a particular
interplay of the pseudo-phonon and pseudo-dilaton modes.
We set the stage for further studies which can naturally follow two directions. The first one
is mainly holographic and aims at solving completely the model at hand,
obtaining therefore direct information on the structure and content of the correlators
that have been only partially constrained by the present analysis. 
The second direction
consists of a systematic 
effective field theory program
based on the explicit realization  of the Ward identity structure unveiled in this work.

The paper is structured as follows. In Section \ref{sec:QFT} we derive in QFT the Ward identities of 
translational and scaling symmetry breaking by a complex scalar operator
that is modulated along one spatial direction.
In Section \ref{olo} we study a holographic 
model realizing the Ward identities found in Section \ref{sec:QFT}. This requires a careful 
renormalization process and proper treatment of bulk gauge invariant quantities.
In Section \ref{gene} we discuss the characteristics of the low-energy modes and 
explore 
generic Ans\"atze for the relevant two-point functions in view of constructing
effective field theories. We conclude in Section~\ref{Discu} taking stock of the results we obtained
and pointing to some important open directions. 

\section{Ward identities}
\label{sec:QFT}
In this section we recall, and quickly re-derive, the Ward identities for one- and two-point functions 
when translations and dilatations are broken, either spontaneously or explicitly.

Consider the generating functional $W[h_{\mu\nu},\phi]$ of connected 
correlators of a $d$-di\-men\-sio\-nal 
QFT defined on Minkowski spacetime.
The field $h_{\mu\nu}$ is coupled to the energy-momentum tensor
while $\phi$ is coupled to a scalar operator $O$. They can be respectively considered as  
small fluctuations upon the background values $ \eta_{\mu\nu}$ and $\bar \Phi$, 
the latter being thus a physical coupling for the operator $O$:
\begin{equation}
S_{QFT} \supset - \frac{1}{2}\int d^dx\ \left(\bar\Phi^*\, O + \bar\Phi\, O^* \right)\ ,
\end{equation}  
which possibly breaks translation invariance (if $\partial_\mu \bar{\Phi}\neq 0$) and scale invariance 
(if the operator $O$ has naive dimension $\Delta_O\neq d$).
Let us denote
\begin{equation}\label{BuBo}
 S_\mathrm{source} = \int d^dx \left[ \frac{1}{2}\, h_{\mu\nu} T^{\mu\nu}  
 - \frac{1}{2} \left(\phi^*\, O + \phi\, O^* \right) \right]\ ,
\end{equation}
where the numerical coefficients are conventional and chosen in view of later convenience. 
The partition function of the system ${\cal Z}$ and the generating functional $W$ are defined as follows
\begin{equation}
{\cal Z} = e^{iW} = \int d \mu\ e^{iS_{QFT}+iS_\mathrm{source}}  \ , 
\label{eq:partfunc}
\end{equation}
where $d \mu$ represents schematically the path integral integration measure. 
Consequently we have
\begin{equation}
\langle T^{\mu\nu}(x) X(x') \rangle =-  2i \frac{\delta}{\delta h_{\mu\nu} (x)} \langle X(x')  \rangle\ ,
\end{equation}
\begin{equation}
\langle O(x)X(x')  \rangle =  2i \frac{\delta}{\delta \phi^* (x)} \langle X(x')  \rangle\ ,\quad
\langle O^*(x) X(x') \rangle =  2i \frac{\delta}{\delta \phi (x)} \langle X(x')  \rangle\ , \label{dictO}
\end{equation}
where $X$ can be replaced by a string of operators at different positions, or by the identity, in which case 
$\langle \mathbbm{1}\rangle \equiv {\cal Z}$.

\subsection{Diffeomorphisms}

Let us now show explicitly how the invariance of the partition function \eqref{eq:partfunc} 
under diffeomorphisms yields the Ward identities
for translations.
The external field fluctuations transform under the infinitesimal 
diffeomorphisms generated by the parameter $\xi^\mu$ 
according to their tensorial structure. Specifically
\begin{subequations}
\label{diffe}
\begin{align}
   \delta_\xi h_{\mu\nu} &=
  \partial_{\mu} \xi_{\nu} + \partial_{\nu} \xi_{\mu} \ , \\
   \delta_\xi \phi&\equiv \delta_\xi \Phi  = \xi_\alpha\, \partial^\alpha \Phi= \xi_\alpha\, \partial^\alpha \bar \Phi 
   + \xi_\alpha\, \partial^\alpha \phi \ .
\end{align}
\end{subequations}
Notice that we did not consider $\mathcal{O}(h)$ corrections to the first equation, while in the second one
we have retained $\mathcal{O}(\phi)$ ones. The reason is that the latter are relevant for the correlators we compute,
while the former are not. We have also  assigned the variation of the  whole field to the fluctuation, effectively taking the background field as invariant, \emph{i.e.}
$\delta_\xi \bar \Phi = 0$.

Using standard techniques, one can derive the following identity for one-point functions
\begin{equation}\label{1WIt}
  \langle \partial_\mu  T^{\mu \nu} \rangle =
  - \frac{1}{2}(\partial^\nu  \Phi^*) \langle O \rangle 
  - \frac{1}{2}(\partial^\nu  \Phi) \langle O^*\rangle \ ,
\end{equation}
which is valid at sources on. Setting the sources to zero is simply equivalent to replacing $\Phi$ with $\bar \Phi$. 

Before computing the two-point functions, we have to implement the shift (as pointed out, for instance, in \cite{Bianchi:2001kw}) of $ T^{\mu \nu}$ to split physical and external sources:
\begin{equation}
  \langle  T^{\mu \nu} \rangle =
   \langle  T^{\mu \nu} \rangle_{QFT}
  - \frac{1}{2} \phi^*\ \eta^{\mu\nu} \langle O \rangle 
  - \frac{1}{2} \phi\ \eta^{\mu\nu}\langle O^*\rangle \ , \label{eq:BFS}
\end{equation}
where $ T^{\mu \nu}_{QFT}$ is the physical energy-momentum tensor, while $ T^{\mu \nu}$ is the one
that takes into account also the part of the action with the external sources, appearing in the
1-pt Ward identity at sources on.
Having done this redefinition, one can finally compute the two-point Ward identity
\begin{equation}\label{2WIt}
 \begin{split}
 & \langle \partial_\mu T^{\mu \nu}(x) O(x') \rangle_{QFT} = 
 i\delta^d(x-x') \partial^{\nu} \langle O(x) \rangle \\
 &\qquad \qquad \qquad - \frac{1}{2} (\partial^\nu \bar \Phi^*(x))\  \langle O(x)O(x') \rangle 
 - \frac{1}{2} (\partial^\nu \bar \Phi(x))\ \langle O^*(x)O(x')\rangle \ ,
 \end{split}
\end{equation}
where the sources have been set to zero.
Note that the shift \eqref{eq:BFS} is crucial in order to obtain the correct contact term.

\subsection{Local scaling transformations}
\label{scalingqft}

The derivation of the Ward identities for scaling transformations is similar to the one 
for translations. The field variations of interest are
\begin{subequations}
\label{scaling}
\begin{align}
 \delta_\beta h_{\mu\nu} &= - 2 \beta\, \eta_{\mu \nu}\ ,\\
  \delta_\beta \phi&=  (d - \Delta_O)\beta\,  \Phi  \ ,
\end{align}
\end{subequations}
where $\beta$ is generically spacetime dependent, $\Delta_O$ is the dimension of the operator $O$,
and we neglect higher order terms.

The one-point Ward identity reads
\begin{equation}\label{1ska}
 \langle T\rangle = -\frac{d - \Delta_O}{2}\  \Phi^*  \langle O\rangle
                    - \frac{d - \Delta_O}{2}\  \Phi  \langle O^*\rangle\ ,                   
\end{equation}
where we can set the external fields to their background values. 
We have defined $T=\eta_{\mu\nu}T^{\mu\nu}$ as customary.

Also in this case, before computing the two-point functions, we need to take into account the redefinition of the energy-momentum tensor \eqref{eq:BFS}, namely we need to consider:
\be
T_{\text{QFT}} = T
+ \frac{d}{2} ( \phi^*\ O +  \phi\ O^* )\,.
\label{eq:BFSska}
\ee
Finally, the resulting 2-pt Ward Identity is
\begin{equation}\label{2ska}
 \begin{split}
 &\langle T(x) O(x') \rangle_{QFT} =\ i\delta^d(x-x') \Delta_O \langle O(x)\rangle\\
  & \qquad \qquad \qquad-\frac{d-\Delta_O}{2}\ \bar \Phi^*(x) \langle O(x) O(x')\rangle 
 -\frac{d-\Delta_O}{2}\ \bar \Phi(x) \langle O^*(x) O(x')\rangle\ .
 \end{split}
\end{equation}

Finally, let us briefly mention the infinitesimal local Lorentz transformations.
Since the sources are scalar, the corresponding 
one- and two-point Ward identities are
\begin{equation}
\langle T^{\left[\mu\nu \right]}(x) \rangle=0 \ , \qquad 
\langle T^{\left[\mu\nu \right]}(x) O(x') \rangle=0 \ .
\end{equation}

The Ward identities derived in this section are crucial for the analysis in Section~\ref{gene}
where we discuss what kind of energy-momentum tensor can realize such Ward identities, and
in particular which gapless (or nearly gapless) modes should appear in the correlator 
$\langle  T^{\mu \nu}(x)\, O(x') \rangle_{QFT}$ in order to satisfy them.
Before that, in the next section 
we turn to holography and present a model that realizes the Ward identities we have just derived.
The holographic model provides also extra input for the analysis of Section~\ref{gene}.

\section{A holographic model}
\label{olo}
In this section we present a holographic realization of translation and scale symmetry breaking. 
Our goal is twofold: first, to realize the Ward identities computed in the previous section; 
second, to obtain useful insight into the structure of the correlators which are sensitive to the physics of the breaking of
translational and scaling symmetry. Such insight will be put to use in the QFT analysis of Section~\ref{gene}. To this purposes, we analyze the holographic Q-lattice \cite{Donos:2013eha,Donos:2014uba}, 
where the symmetry breaking (in particular that of translations) is encoded in a dual gravitational model with a 
spatially-oscillating complex scalar field in an asymptotically $AdS$ spacetime.
More precisely, let us consider the action of a massive complex scalar field $\Phi$ of mass $m$ in $d+1=4$  dimensional spacetime:
\begin{equation}\label{holoac}
 S =S_0+S_{GH}+S_{\text{c.t.}} \ , \qquad 
 S_0 = \int d^4 x\ \sqrt{-g} \left( R - 2\Lambda - \partial_M \Phi\, \partial^M \Phi^*  - m^2 \Phi \Phi^* \right) \ ,
\end{equation}
where the capital Latin indices run over the bulk coordinates $(t,x,y,z)$. We henceforth fix the cosmological constant $\Lambda$ to $-3$.
This model is dual to a CFT which has a complex scalar operator%
\footnote{As usual in the literature using such holographic Q-lattices, we will just assume the existence of a regime where a gravitational theory features a globally charged scalar.}
with
scaling dimension $\Delta = \frac{d}{2}\pm\sqrt{\frac{d^2}{4}+m^2}$.
We consider for definiteness $m^2=-2$ so that $\Delta=2$, noting that the specific value of the 
scaling dimension of the operator is nowhere crucial to the results obtained in this section.
The Gibbons-Hawking term, $S_{GH}$ in \eqref{holoac}, is the boundary term required to have a well-defined 
variational problem, and it is given by
\begin{equation}
S_{GH} = \int d^3 x \ \sqrt{-\gamma}\ 2 K = \int d^3 x \ \sqrt{-\gamma}\ 2 g^{MN} \nabla_M n_N\ ,\label{SGH}
\end{equation}
where $n_M$ is the unit vector orthogonal to the boundary, and $\gamma$ the determinant of the metric with the 
direction along $n_M$ excluded. The counterterm part of the action \eqref{holoac}, $S_{\text{c.t.}}$, 
takes into account the counterterms needed to holographically renormalize the action, which, for the model at hand, are given by: 
\begin{equation}\label{count}
S_{\text{c.t.}}=\int d^3 x\ \sqrt{-\gamma} \left(4+R\left[\gamma\right]+\Phi\Phi^*\right) \ .
\end{equation} 
The equations of motion can be written as follows:
\begin{subequations}
\label{eom}
\begin{eqnarray}\label{eq:eqeins}
 & R_{MN} - \frac{1}{2} R g_{MN} + \Lambda g_{MN} = - \frac{1}{2} g_{MN} \left( \partial_A \Phi \partial^A \Phi^* 
 + m^2 \Phi\Phi^* \right)
 + \partial_{(M} \Phi \partial_{N)}\Phi^* \ , \\
& \partial_M \left(\sqrt{-g}\, g^{MN} \partial_N \Phi\right) - \sqrt{-g}\, m^2 \Phi = 0 \ , \label{eqscal}
\end{eqnarray}
\end{subequations}
plus an equation analogous to \eqref{eqscal} for $\Phi^*$.

We  are  interested  in analyzing the zero-temperature case with the scalar operator breaking the translation 
symmetry in one of the spatial directions by means of an oscillating phase $e^{i \,k\cdot x}$, where
the wave-vector is spatial and taken along the $x$ direction, $k=(0,k_x,0)$.
Accordingly, we consider the following Ansatz
\begin{equation}
 \begin{split}\label{assa}
  ds^2 & = \bar g_{MN}dx^M dx^N = -T(z) dt^2+X(z) dx^2+ Y(z) dy^2+Z(z) dz^2 \ ,\\
  \Phi(x,z)&= e^{ik\cdot x} \varphi(z) \ .
 \end{split}
\end{equation}
Standard AdS spacetime in the Poincar\'e patch corresponds to $T(z)=X(z)=Y(Z)=Z(z)=\frac{1}{z^2}$ where $z$ is the radial coordinate.
The bar over $\bar g_{MN}$ indicates that it is the background (we reserve the unbarred symbol for
the whole background-plus-fluctuation field).

The analysis of the Ward identities described in the following  relies only on the UV properties of the model. 
It is therefore sufficient to consider an asymptotic solution near the conformal boundary ($z=0$). 
The expansion of the bulk background fields as a power series in $z$ is given by
\begin{align}\label{ansatz1}
 \varphi(z) &= \varphi_1 z + \varphi_2 z^2 +... \ , \\ \nonumber
 \bar g_{MM}(z) &= z^{-2} \left(\eta_{MM} + g_{1MM}\ z + g_{2MM}\ z^2 + g_{3MM}\ z^3 +...\right) \ .
\end{align}
Relying on the scaling properties of the model, the coefficient of the leading term of the metric components has been set to one.
We take $\varphi(z)$ to be a real field. 
Solving order by order the equations of motion \eqref{eom} we have
\begin{align} \nonumber
 \varphi(z) &=  \varphi_1 z + \varphi_2 z^2 + \mathcal{O}(z^3) \ ,\\ \nonumber
 z^2 T(z) &= 1 -\frac{\varphi_1^2}{4} z^2 + T_3 z^3 + {\cal O}(z^4) \ ,\\ \label{backsol}
 z^2 X(z) &= 1 -\frac{\varphi_1^2}{4} z^2 + X_3 z^3 + {\cal O}(z^4) \ ,\\ \nonumber
 z^2 Y(z) &= 1 -\frac{\varphi_1^2}{4} z^2 + Y_3 z^3 + {\cal O}(z^4) \ ,\\ \nonumber
 z^2 Z(z) &= 1 - \left(T_3+X_3+Y_3+\frac{4}{3}\varphi_1\varphi_2 \right) z^3 + {\cal O}(z^4) \ .
\end{align}
We adopted a convenient way to fix the remaining
gauge arbitrariness related to radial diffeomorphisms by setting $g_{1zz}=g_{2zz}=0$, and we further impose the condition
\begin{equation}
 g_{3zz} = 0\qquad \Leftrightarrow \qquad T_3+X_3+Y_3+\frac{4}{3}\varphi_1\varphi_2=0\ .
\end{equation}
Note that in \eqref{backsol} the $z$-linear terms in the metric components vanish because of the EOMs. 

It is important to note that the background metric does not depend on the boundary coordinates.
 This is a crucial simplification particular to the Q-lattice ansatz that will be pivotal for
the analysis in what follows. Indeed, it 
can also be shown that the ${\cal O}(z^4)$ terms will depend on $k$ only through $k^2$.

\subsection{Fluctuations}

In order to compute the one- and two-point Ward identities we need to consider the fluctuations of the fields 
of the model around the background solution \eqref{backsol}. More specifically, we are interested in computing 
the action \eqref{holoac} expanded up to second order in the fluctuations. Consider the fluctuations
\begin{equation}\label{flu}
g_{MN} = \bar g_{MN} + h_{MN} \ , \qquad
\Phi = \bar \Phi + \phi \ ,
\end{equation}
where $\bar g_{MN}$ and $\bar \Phi$ are the background values displayed in \eqref{backsol}.

The equations of motion \eqref{eom} expanded at the linear order in the fluctuations \eqref{flu} read
\begin{equation}\label{scaflu}
 -h^{MA} \nabla_M \nabla_A \bar\Phi
 -\nabla_M \bar \Phi \nabla_A h^{MA} 
 +\frac{1}{2} \nabla^M \bar \Phi \nabla_M h^A_{\ A}
 +\bar g^{MA} \nabla_A \nabla_M \phi
 -m^2 \phi = 0 \ ,
\end{equation}
\begin{equation}\label{nosflu}
 \begin{split}
&\frac{1}{2} \nabla_B\nabla_A h_C^{\ C}
 -\frac{1}{2} \nabla_C\nabla_A h_B^{\ C}
 -\frac{1}{2} \nabla_C \nabla_B h_A^{\ C}
 +\frac{1}{2} \nabla_C \nabla^C h_{AB}
 +\Lambda\, h_{AB}\\
&+\frac{1}{2} m^2 \bar \Phi \bar \Phi^* h_{AB}
 +\frac{1}{2} m^2 \bar \Phi^*\, \phi\, \bar g_{AB}
 +\frac{1}{2} m^2 \bar \Phi\, \phi^*\, \bar g_{AB}\\
&+\frac{1}{2} \nabla_A \bar \Phi^* \nabla_B \phi
 +\frac{1}{2} \nabla_B \bar \Phi^*  \nabla_A \phi
 +\frac{1}{2} \nabla_A \bar \Phi \nabla_B \phi^*
 +\frac{1}{2} \nabla_B \bar \Phi  \nabla_A \phi^*
 = 0 \ .
 \end{split}
\end{equation}
The terms of the bulk action $S_0$ \eqref{holoac} that are linear and quadratic
in the fluctuations $\phi$ and $h_{MN}$ 
reduce to purely boundary terms upon using respectively 
the equations of motion of the background and of the fluctuations.
The term linear in the fluctuation gives
\begin{equation}
S^{(1)}_0 = \int_{z=\epsilon} d^3 x\  \sqrt{-\bar g} \Big[\nabla_N h^{zN}-\nabla^z {h^N}_N 
- \partial^z\bar\Phi\, \phi^*
- \partial^z\bar\Phi^*\, \phi\Big] \ , \label{onsh1}
\end{equation}
while the quadratic term takes the form
\begin{align}
S^{(2)}_0 = \int_{z=\epsilon} d^3 &x\ \sqrt{-\bar g} \Big[
-\frac14 {h^N}_N \nabla^z {h^M}_M
-h^{zM}\nabla^N h_{NM}
+\frac34 h_{MN}\nabla^z h^{MN}\nn\\
&-\frac12 h_{MN}\nabla^M h^{zN}  +\frac14 {h^N}_N \nabla_M h^{M z}
+\frac34 h^{zM}\nabla_M{h^N}_N 
\label{onsh2} \nn\\
&+\frac12 h^{zN}(\phi \partial_N \bar \Phi^*
+\phi^* \partial_N \bar\Phi) -\frac14 {h^N}_N (\phi \partial^z \bar \Phi^*
+ \phi^* \partial^z \bar \Phi) \nn \\
&-\frac12 \phi \partial^z \phi^* -\frac12 \phi^*  \partial^z \phi\Big] \ , 
\end{align}
where $\epsilon$ is a UV cut-off which is set to zero at the end of the 
renormalization process.
To renormalize $S_0$, we need to add to \eqref{onsh1} and \eqref{onsh2} 
the boundary terms \eqref{SGH} and \eqref{count} computed at $z=\epsilon$ and eventually 
to perform the $\epsilon \rightarrow 0$ limit. To this end, it is again convenient
to adopt the radial gauge
\begin{equation}\label{radiga}
h_{z M}=0 \ .
\end{equation}
In this gauge the equations of motion \eqref{scaflu} and \eqref{nosflu} reduce to a set 
of seven dynamical equations plus four constraints coming from the equations for $h_{z M}$, 
and the Gibbons-Hawking term \eqref{SGH} reads
\begin{equation}\label{SGHr}
S_{GH}= \int_{z=\epsilon} d^3x\ \sqrt{-\gamma}\ g^{\mu\nu}\partial^z g_{\mu\nu}\ ,
\end{equation}
where the indices $\mu,\nu$ label the boundary coordinates.

Let us first focus on the linear part of the action. Expanding \eqref{SGHr} to linear order we obtain:
\begin{equation}
S_0^{(1)}+S_{GH}^{(1)}= \int_{z=\epsilon} d^3x\ \sqrt{-\bar g}
\left[ \frac12h_{\mu\nu} \left( \partial^z \bar g^{\mu\nu}
-\bar g^{\mu\nu} \bar g_{\alpha\beta}\partial^z \bar g^{\alpha\beta}\right)
-\partial^z \bar \Phi \phi^* 
-\partial^z \bar \Phi^* \phi\right] \ .
\end{equation}
We now write the expansions for all the fields, both background and fluctuations, near $z=0$ in the following way
(the expansion of the background metric is just a repetition of \eqref{ansatz1} with boundary indices only):
\begin{align}\label{ansaflu}
\bar g_{\mu\nu} &= \frac1{z^2}\left( \eta_{\mu\nu} + g_{2\mu\nu}z^2 + g_{3\mu\nu}z^3+\dots\right) \ , \\ \nn
h_{\mu\nu} &= \frac1{z^2}\left( h_{0\mu\nu} + h_{2\mu\nu}z^2 + h_{3\mu\nu}z^3+\dots\right)  \ ,\\ \nn
\bar \Phi & = \bar \Phi_1 z + \bar \Phi_2 z^2 + ... = e^{i k x}\left(\varphi_1 z + \varphi_2 z^2 +\dots \right) \ ,\\ \nn
\phi & = \phi_1 z + \phi_2 z^2 + ... = e^{i k x}\left(\hat\varphi_1 z + \hat\varphi_2 z^2 +\dots\right) \ . 
\end{align}
The coefficients of the fluctuations $h_{0\mu\nu}$, $h_{2\mu\nu}$, $h_{3\mu\nu}$, 
$\hat\varphi_1$ and $\hat\varphi_2$ are generic functions of $x^\mu$.

Adding all the counterterms \eqref{count} expanded to linear order in the fluctuations
and performing the $\epsilon \rightarrow 0$ limit, we get
\begin{equation}\label{line11}
S^{(1)}_\mathrm{ren} = \int_{z=0} d^3x
\left[ -\frac32 g_{3\mu\nu}h^{\mu\nu}_0
+\frac34 g_{3\mu}^{\ \ \mu}h_{0\nu}^{\ \ \nu}
-\varphi_2\left(\hat\varphi_1 +\hat\varphi_1^*\right)\right] \ .
\end{equation}

We now turn to the second order terms in the action. 
Expanding \eqref{SGHr} to second order in the fluctuations, we obtain
\begin{align}
S^{(2)}+S_{GH}^{(2)}=  \int_{z=\epsilon}& d^3x\ \sqrt{-\bar g}\left[
\frac14 {h^\lambda}_\lambda \left( \bar g^{\mu\nu}\partial^z h_{\mu\nu}
+\frac32 h_{\mu\nu}\partial^z \bar g^{\mu\nu}\right) 
-\frac14 h^{\mu\nu}\partial^z h_{\mu\nu}\right. \nn \\  
-&\frac12 h_{\mu\lambda}{h^\lambda}_\nu \partial^z \bar g^{\mu\nu}  
-\frac14\left( h^{\lambda \omega}h_{\lambda \omega}
-\frac12{h^\lambda}_\lambda{h^\omega}_\omega\right) \bar g^{\mu\nu}\partial^z \bar g_{\mu\nu}\nn\\
&\left. -\frac14{h^\mu}_\mu(\phi\partial^z\bar\Phi^*
+\phi^*\partial^z\bar\Phi)
-\frac12 (\phi\partial^z \phi^* + \phi^*\partial^z \phi)\right] \ . 
\end{align}
We now add the counterterms \eqref{count} expanded to the second order in the fluctuations and, 
having expanded the fields near $z=0$ as in \eqref{ansaflu},  we send $\epsilon \rightarrow 0$. 
Divergent terms arise at $\mathcal{O}(z^{-3})$ and  $\mathcal{O}(z^{-1})$ but they all cancel. 
For the latter, it is crucial to substitute the following expression in the on-shell action, 
obtained through the $\mu\nu$ equation of the set \eqref{nosflu}:
\begin{align}
h_{2\mu\nu}=& \frac12(\partial_\lambda\partial^\lambda h_{0\mu\nu}
+\partial_\mu\partial_\nu h_{0\lambda}^{\ \ \lambda}
-\partial_\lambda\partial_\mu h_{0\nu}^{\ \ \lambda}
-\partial_\lambda\partial_\nu h_{0\mu}^{\ \ \lambda})
+\frac14\eta_{\mu\nu}(\partial_\lambda \partial_\omega h_0^{\ \lambda\omega}
-\partial_\lambda \partial^\lambda h_{0\omega}^{\ \ \omega}) \nn\\
&-\frac14 \varphi_1^2h_{0\mu\nu}
-\frac14 \varphi_1 \eta_{\mu\nu}(\hat \varphi_1+\hat\varphi_1^*)\ .
\end{align}
Note that we have substituted the value $g_{2\mu\nu}=-\frac14 \varphi_1^2\eta_{\mu\nu}$ 
according to the background solution \eqref{backsol}. The counterterm proportional to $R$ 
is needed to cancel the remaining $\mathcal{O}(z^{-1})$ divergences, but does not affect the finite terms.

We eventually get
\begin{align}\label{second11}
S^{(2)}_\mathrm{ren} =  \int d^3x\ &\left[ -\frac34
h_{0\mu\nu}h_3^{\ \mu\nu}
+\frac34 h_{0\mu}^{\ \ \mu}h_{3\nu}^{\ \ \nu} 
+ \frac32 g_{3\mu\nu}\left( h_{0\lambda}^{\ \ \mu}h_0^{\ \nu\lambda}
-\frac34 h_0^{\ \mu\nu}h_{0\lambda}^{\ \ \lambda}\right) \right. \nn \\
&\quad -\frac14\left( 3g_{3\mu}^{\ \ \mu}+2\varphi_1\varphi_2\right) \left( h_{0\nu\lambda}h_0^{\ \nu\lambda}
-\frac12  h_{0\nu}^{\ \ \nu}h_{0\lambda}^{\ \ \lambda}\right) \nn\\
&\quad \left. +\frac14 h_{0\mu}^{\ \ \mu}\varphi_1(\hat\varphi_2+\hat\varphi_2^*)
-\frac12(\hat\varphi_1\hat\varphi_2^*+\hat\varphi_1^*\hat\varphi_2)\right] \ .
\end{align}
According to the holographic dictionary, this expression serves as a generating 
functional $W$ when it is written in terms of the sources only.
In this setup the sources are $h_{0\mu\nu}$ and $\hat\varphi_1$. Notice that
in \eqref{second11}  the dependence on these fields is still implicit 
in $h_{3 \mu\nu}$ and $\hat\varphi_2$.

We first use the constraints, coming from the $zz$ and $z\mu$ Einstein equations \eqref{nosflu}, in 
order to express those parts of $h_{3\mu\nu}$ which are not genuinely independent from the sources. 
The constraint equations are 
\begin{equation}\label{cos}
 2 [ \varphi_1(\hat \varphi_2 + \hat \varphi_2^*) + \varphi_2(\hat \varphi_1  + \hat  \varphi_1^*)]
 -3 h_{0\mu\nu} g_3^{\ \mu\nu} +3 h_{3\mu}^{\ \ \mu}  =0\ ,
\end{equation}
and
\begin{align}\label{cosa}
\partial^\nu h_{3\nu\mu}= & \ g_{3\mu\nu}\partial_\lambda h_0^{\ \nu\lambda}
+\frac12 g_{3 \nu\lambda}\partial_\mu h_0^{\ \nu\lambda}
-\frac12 g_{3\mu\nu} \partial^\nu h_{0\lambda}^{\ \ \lambda} \nn \\
& +\frac13 ik_\mu \left[ \varphi_2 (\hat\varphi_1-\hat\varphi_1^*)-\varphi_1 (\hat\varphi_2-\hat\varphi_2^*)\right] 
-\frac13 \varphi_1 \partial_\mu (\hat\varphi_2+\hat\varphi_2^*)\ .
\end{align}
In order to implement the second constraint in \eqref{second11}, we need to split the 
source $h_{0\mu\nu}$ into its irreducible components:
\begin{equation}\label{dofa}
  h_{0\mu\nu} =  h_{0\mu\nu}^{(tt)} + \partial_{(\mu}  h^{(t)}_{0\nu)} 
 + \eta_{\mu\nu}  h_0 + \frac{\partial_\mu \partial_\nu}{\Box}  H_0\ ,
\end{equation}
where $\partial^\mu h_{0\mu\nu}^{(tt)}=0$, $h_{0\mu}^{(tt)\mu}=0$ and $\partial^\mu h^{(t)}_{0\mu}=0$.
 Notice that this is the usual splitting of the metric into irreducible representations
of the Lorentz group, yet can also be seen as a mere introduction of a basis. 

As a result of the decomposition above
we obtain
\begin{align}
S^{(2)}_\mathrm{ren} \supset  \int d^3x\ &\left\{-\frac{3}{4}h_{0\mu\nu}^{(tt)}\, h_3^{(tt)\mu \nu}+ \frac i4k^\mu  \left( h^{(t)}_{0\mu}+ \frac{ \partial_\mu}{\Box}  H_0\right) \left[ \varphi_2 (\hat\varphi_1-\hat\varphi_1^*)-\varphi_1 (\hat\varphi_2-\hat\varphi_2^*)\right]\right. \nn \\
&\quad -\frac14 \varphi_1h_0 (\hat \varphi_2 + \hat \varphi_2^*)
- \varphi_2h_0(\hat \varphi_1  + \hat  \varphi_1^*)
-\frac12  \varphi_2H_0(\hat \varphi_1  + \hat  \varphi_1^*) \nn\\
&\quad \left.
-\frac12(\hat\varphi_1\hat\varphi_2^*+\hat\varphi_1^*\hat\varphi_2)\right\} \ ,\label{srenint}
\end{align}
 where we have displayed only the terms depending on $\hat\varphi$ explicitly or implicitly since 
only those terms contribute to the mixed correlators of the scalar operators and the energy-momentum tensor we are
interested in. Notice that in the first line we have kept the term containing the transverse traceless part of $h_{3\mu\nu}$, which might depend
implicitly on $\hat\varphi$.

The dependence of the VEVs, $h_{3\mu\nu}^{(tt)}$, $\hat\varphi_2$ and $\hat\varphi_2^*$,
on the sources must respect gauge invariance.
While $h_{\mu\nu}^{(tt)}$ is gauge invariant by itself, 
for the scalar fields and the remaining components 
of the metric we have to write combinations invariant under
\begin{align}\label{trlt}
 \delta  h_{\mu\nu} = \partial_\mu \xi^{(t)}_\nu + \partial_\nu  \xi^{(t)}_\mu 
 + \frac{\partial_\mu \partial_\nu}{\Box} \chi - 2 \beta\, \eta_{\mu\nu} \ ,
\end{align}
\begin{equation}\label{skatra}
 \delta \phi =  \xi_\mu^{(t)} \partial^\mu \bar\Phi 
 + \frac{1}{2} \left(\frac{\partial_\mu}{\Box} \chi \right) \partial^\mu \bar\Phi\ +\beta \,z\, \partial_z \bar \Phi\ ,
\end{equation}
where we have split $\xi_\mu = \xi^{(t)}_\mu +\frac12 \frac{\partial_\mu }{\Box} \chi$, 
with $\partial^\mu \xi^{(t)}_\mu=0$.
These transformations correspond to the diffeomorphisms and local scale transformations
defined in \eqref{diffe} and \eqref{scaling}, respectively. Using the expansions \eqref{ansaflu}, and the
decomposition \eqref{dofa}, we arrive to
\begin{align}
\delta h_{0\mu}^{(t)}&=2\xi^{(t)}_\mu\ ,\label{tfht} \\
\delta H_{0}&=\chi\ , \\
\delta h_{0}& = -2\beta\ , \\
\delta \hat\varphi_1&=ik^\mu \left(\xi^{(t)}_\mu
+ \frac12 \frac {\partial_\mu}{\Box}\chi\right)\varphi_1+\beta \varphi_1\ ,
\label{tfphi1} \\
\delta \hat\varphi_2&=ik^\mu \left(\xi^{(t)}_\mu
+ \frac12 \frac {\partial_\mu}{\Box}\chi\right)\varphi_2+2\beta \varphi_2\ .
\end{align}
The invariant combinations are thus
\begin{align}\label{spietta1}
 \spy_1  &= \hat \varphi_1 -i k^\mu \frac12 \left(\delta h_{0\mu}^{(t)}+\frac {\partial_\mu}{\Box}H_0\right) \varphi_1
              +\frac12 h_0 \varphi_1\ ,\\
\label{spietta}
 \spy_2   &=\hat \varphi_2 -i k^\mu \frac12 \left(\delta h_{0\mu}^{(t)}+\frac {\partial_\mu}{\Box}H_0\right) \varphi_2
               + h_0 \varphi_2\ .
\end{align}
We now express all the components of the metric sources introduced in \eqref{dofa} in
terms of projectors acting on $h_{0\mu\nu}$:
\begin{align}
 h_0 &= \frac{1}{2} \left( \eta^{\mu\nu} - \frac{\partial^\mu \partial^\nu}{\Box} \right) h_{0\mu\nu} \ ,\\
 H_0 &= - \frac{1}{2} \left( \eta^{\mu\nu} - 3 \frac{\partial^\mu \partial^\nu}{\Box} \right)  h_{0\mu\nu} \ , \\
 h^{(t)}_{0\mu} &=2\left(\frac{\partial^{\nu}}{\Box} \delta_\mu^{\kappa} -
                \partial_\mu \frac{\partial^\nu \partial^\kappa}{\Box^2} \right)  h_{0\nu\kappa} \ , \\
h_{0\mu\nu}^{(tt)} &= \mathcal{T}^{\alpha \beta}_{\mu \nu}  h_{0\alpha\beta}        \ ,      
\end{align}
with 
\begin{align}\label{defprojtt}
\mathcal{T}_{\mu \nu}^{\alpha \beta}&= \delta_{\mu}^{\alpha} \delta_{\nu}^{\beta}
-2 \partial_{(\mu} \left(\frac{\partial^{\alpha}}{\Box}\delta^{\beta}_{\nu)}
-\partial_{\nu)}\frac{\partial^{\alpha}\partial^{\beta}}{\Box^2}\right)\nn\\
&\quad-\frac{1}{2} \eta_{\mu \nu} \left( \eta^{\alpha \beta}- \frac{\partial^{\alpha} \partial^{\beta}}{\Box}\right)+\frac{1}{2}\frac{\partial_{\mu} \partial_{\nu}}{\Box}\left(\eta^{\alpha \beta}-3 \frac{\partial^{\alpha} \partial^{\beta}}{\Box}\right) \ .
\end{align} 
Using these expressions, the gauge invariant quantities become
\begin{align*}
\spy_1=& \hat \varphi_1 -i k^\mu\,{\cal D}^{\alpha\beta}_\mu\, h_{0\alpha\beta}\, \varphi_1
+ \frac{1}{4}\, {\cal P}^{\alpha\beta}\, h_{0\alpha\beta}\, \varphi_1\ , \\
\spy_2=& \hat \varphi_2 -i k^\mu\,{\cal D}^{\alpha\beta}_\mu\, h_{0\alpha\beta}\, \varphi_2
+ \frac{1}{2}\, {\cal P}^{\alpha\beta}\, h_{0\alpha\beta}\, \varphi_2\ ,
\end{align*} 
where the projectors ${\cal D}^{\alpha\beta}_\mu$ and ${\cal P}^{\alpha\beta}$ 
are defined as follows
\begin{equation}
 {\cal D}^{\alpha\beta}_\mu=
\frac{\delta_\mu^\alpha\,\partial^\beta+\delta_\mu^\beta\,\partial^\alpha}{2\Box}
- \frac{\eta^{\alpha\beta}}{4} \frac{\partial_\mu}{\Box}
- \frac{\partial^\alpha\partial^\beta\partial_\mu}{4\Box^2}\,, \qquad{\cal P}^{\alpha\beta}=
\left(\eta^{\alpha\beta}-{\partial^\alpha\partial^\beta\over\Box}\right)\,. \label{diffops}
\end{equation}
Gauge invariance implies that
\begin{align}
\spy_2(x)=&\int d^3x' \left[f(x-x')\,\spy_1(x')+g(x-x')\,\spy_1^*(x')+c(x-x')\,k^{\mu} k^{\nu} h_{0\mu\nu}^{(tt)}(x') \right]
\nn\\
\equiv&\;f(\partial)\spy_1(x) + g(\partial)\spy_1^*(x)+c(\partial)\, k^{\mu} k^{\nu}h_{0\mu\nu}^{(tt)}(x)\ ,\\
\label{ttmetricgi}
h_{3\mu\nu}^{(tt)}=&\int d^3x' \left[e(x-x') h_{0\mu\nu}^{(tt)}(x')+\right.\nn\\
&\left.\frac{2}{3}h(x-x')k_{\alpha}k_{\beta}\mathcal{T}^{\alpha \beta}_{\mu \nu}\spy_1(x')
+\frac{2}{3}h^*(x-x')k_{\alpha}k_{\beta}\mathcal{T}^{\alpha \beta}_{\mu \nu}\spy^*_1(x') \right]\nn\\
 \equiv&\;e(\partial)h_{0\mu\nu}^{(tt)}(x)
+\frac{2}{3}h(\partial)k_{\alpha}k_{\beta}\mathcal{T}^{\alpha \beta}_{\mu \nu}\spy_1(x)
+\frac{2}{3}h^*(\partial)k_{\alpha}k_{\beta}\mathcal{T}^{\alpha \beta}_{\mu \nu}\spy^*_1(x) \ ,
\end{align}
where we have taken into account that the metric has to be real and we have put a factor $\frac{2}{3}$ in front of the function $h(x-x')$ just for later convenience. 

The quantities $c(x-x')$, $e(x-x')$, $f(x-x')$, $g(x-x')$ and $h(x-x')$, or alternatively the differential operators $c(\partial)$, $e(\partial)$, $f(\partial)$, $g(\partial)$ and $h(\partial)$,
encode the IR information of the system (hence are typically non-local), and 
can be fully determined only by solving the model.

We can finally solve for $\hat \varphi_2$: 
\begin{align}
\hat \varphi_2 = &f(\partial) \hat \varphi_1 
+ g(\partial) \hat \varphi_1^* + i k^\mu \left[ \varphi_2 
-\varphi_1 (f(\partial)-g(\partial))\right] {\cal D}^{\alpha\beta}_\mu h_{0\alpha\beta}\nn \\ 
 &\qquad \qquad - \frac{1}{4}\left[ 2\varphi_2 -\varphi_1 (f(\partial)+g(\partial))\right] {\cal P}^{\alpha\beta}h_{0\alpha\beta}+c(\partial)k^{\mu}k^{\nu} \mathcal{T}_{\mu \nu}^{\alpha \beta} h_{0 \alpha \beta} \ .\label{vevo}
\end{align}
Substituting the expressions \eqref{ttmetricgi} and \eqref{vevo} in \eqref{srenint} (assuming  $f(\partial)$ to be
real\footnote{We could actually relax this assumption and take $f(\partial)$ to be complex, but only its real part will eventually appear in the renormalized action below.}), we get the following terms containing at least one $\hat\varphi_1$ 
\begin{align}
S^{(2)}_\mathrm{ren} \supset  \int d^3x\ &\left\lbrace \frac{}{} ik^\mu\,{\cal D}^{\nu\lambda}_\mu\,  h_{0\nu\lambda}
\left[ \left(\varphi_2 -\varphi_1\,f(\partial)\right)(\hat\varphi_1-\hat\varphi_1^*)
+\varphi_1\left(g^*(\partial)\,\hat\varphi_1-g(\partial)\,\hat\varphi_1^*\right)\right]\right. \nn \\
& \quad- \frac{1}{2} \ k^{\mu} k^{\nu} \mathcal{T}_{\mu \nu}^{\alpha \beta}h_{0 \alpha \beta}\left[\left(c(\partial)+h(\partial) \right)\hat{\varphi}_1+\left(c^*(\partial)+h^*(\partial) \right)\hat{\varphi}_1^* \right] \nn\\
& \quad
-\frac{1}{4}\varphi_1\, {\cal P}^{\mu\nu}\, h_{0\mu\nu}  \left[f(\partial)(\hat \varphi_1  + \hat  \varphi_1^*)
+g(\partial)\,\hat\varphi_1^* + g^*(\partial)\,\hat\varphi_1\right]
\label{srenfin} \\
&\quad \left.
-\frac12  \varphi_2\,  \frac{\partial^\mu \partial^\nu}{\Box}\,  h_{0\mu\nu}\,(\hat \varphi_1  + \hat  \varphi_1^*)
-\hat\varphi_1f(\partial)\hat\varphi_1^* \right. \nn \\  &\quad\left. -\frac12\hat\varphi_1\,g^*(\partial)\,\hat\varphi_1 
-\frac12\hat\varphi_1^*\,g(\partial)\,\hat\varphi_1^*\right\rbrace \ .\nn
\end{align}
We have thus completed the holographic computation of the pieces of the generating functional $W\equiv S^{(1)}_\mathrm{ren}+S^{(2)}_\mathrm{ren}$ needed to find the correlators of interest.

\subsection{Ward identities}
We shall now verify that the generating functional obtained above encodes the Ward identities obtained in the 
QFT analysis
of Section~\ref{sec:QFT}.
This follows from the fact that the generating functional is invariant under the relevant local symmetries,  
diffeomorphisms and local rescalings, acting as (\ref{trlt}, \ref{skatra}).

According to the holographic dictionary, one interprets
\eqref{BuBo} as the bulk to boundary coupling, namely
\begin{equation}\label{holodefbb}
\delta W = \int d^3 x\ \left[\frac{1}{2}\, \langle T^{\mu\nu}\rangle \delta h_{0\mu\nu}
-\frac{1}{2} \left( \delta\phi^*_1 \langle O \rangle 
+\delta\phi_1\langle  O^* \rangle \right)  \right]\ ,
\end{equation}
\emph{i.e.} the leading terms of the bulk fields \eqref{ansaflu} are 
holographically identified with the external sources of the boundary QFT. The one-point functions in the expression above are the ones at sources on. 

By applying functional derivatives with respect to those external sources as in \eqref{dictO}, 
 one obtains the following dictionary. From \eqref{line11} and redefining $O=e^{ikx}O_\varphi$, one gets
\begin{equation}\label{dc1}
 \langle O_\varphi \rangle=2 \varphi_2 \ 
\end{equation}
after setting the sources to zero. From \eqref{srenfin} one gets
\begin{align}
&\langle O_\varphi^*(x)\, O_\varphi(x') \rangle= 4 i f(x-x') \ ,\nn\\ 
&\langle O_\varphi(x)\, O_\varphi(x') \rangle=4 i g(x-x')\ ,\qquad
\langle O_\varphi^*(x)\, O_\varphi^*(x') \rangle=4 i g^*(x-x')\ .
\label{corresca}
\end{align}

The one-point Ward identities are obtained considering the linear on-shell action
$S^{(1)}_{\text{ren}}$ given by \eqref{line11}. It is straightforward to check that it is invariant under both diffeomorphisms and local rescalings. Varying only the metric source produces the following identities
\begin{equation}
\langle \partial_{\mu} T^{\mu \nu} \rangle=0\ , \qquad
\langle T \rangle = -2\varphi_1\,\varphi_2= - \varphi_1 \, \langle O_\varphi \rangle\ ,
\end{equation}
which actually coincide with \eqref{1WIt} and \eqref{1ska} after taking into account that
in the Q-lattice Ansatz \eqref{assa} the field $\varphi$ is real.
Then, at the level of the one-point Ward identities the holographic Q-lattice behaves effectively 
as a translational invariant system.

Let us now focus on the two-point Ward identities, for which we need to consider $S^{(2)}_{\text{ren}}$. 
Notice that $S^{(2)}_{\text{ren}}$ can be shown to be invariant under local transformations upon taking
into account the contribution at second order from the transformation of $S^{(1)}_{\text{ren}}$.
To derive an expression for the two-point functions $\langle \partial_\mu T^{\mu \nu}(x) O_\varphi(x') \rangle$ and
$\langle T(x)O_\varphi(x')\rangle$, one needs to vary the generating functional as in \eqref{holodefbb} for a specific transformation of $h_{0\mu\nu}$, and then take a functional derivative with respect to $\hat{\varphi}_1$. Equating what we expect by definition, with the result from using \eqref{srenfin}, we get
\begin{multline}\label{WIt}
 \langle \partial_\mu T^{\mu \nu}(x)\, O_\varphi(x') \rangle =
  -k^{\nu} \langle O_\varphi \rangle \delta^3(x-x')\\
 -\frac{i}{2} k^{\nu} \varphi_1 \langle O_\varphi^*(x)\, O_\varphi(x') \rangle
 +\frac{i}{2} k^{\nu} \varphi_1 \langle O_\varphi(x)\, O_\varphi(x') \rangle 
 -i \langle O_\varphi\, \rangle \partial^{\nu} \delta^3(x-x')  
 \ ,
\end{multline}
\begin{equation}\label{2ptholodit}
 \langle T(x)\,O_\varphi(x')\rangle = -i \langle O_\varphi \rangle \delta^3(x-x') 
 - \frac{1}{2} \varphi_1  \langle O_\varphi^*(x)\, O_\varphi(x') \rangle
 - \frac{1}{2} \varphi_1  \langle O_\varphi(x)\, O_\varphi(x') \rangle \ ,
\end{equation}
where we have used the holographic dictionary \eqref{dc1} and \eqref{corresca}.
The Ward identities above have to be compared with the ones obtained in the QFT framework, 
namely \eqref{2WIt} and \eqref{2ska} upon implementation of the shift \eqref{eq:BFS}.
As we show explicitly below (see eqs. (\ref{eq:witrhol}, \ref{eq:widilhol})), both sets of Ward identities agree.

We have thus proven that the holographic on-shell action reproduces exactly the Ward identities that we 
computed with standard field theory techniques. 
This may not be a complete surprise, considering that Ward identities are just 
a reflection of the symmetries of the problem, and the fact that the holographic model is built precisely
to encode those symmetries. In other words, once we obtain a generating functional that is invariant under
the relevant symmetries, the Ward identities are guaranteed to follow.

\subsection{Full mixed correlator}

Relying on the holographic model, we can actually obtain an explicit expression for the full mixed correlator
$\langle T_{\mu\nu}\, O_\varphi \rangle$. This expression is used in Section~\ref{gene} to justify a generic Ansatz for 
$\langle T_{\mu\nu}\, O_\varphi \rangle$ which, in turn, is employed to find the gapless or nearly gapless modes related
to the broken symmetries. 

From \eqref{srenfin} one gets
\begin{align}\label{holocorre}
\langle T^{\mu\nu}(x)\, O_\varphi (x')\rangle =
& \,4i\frac{\delta^2 S_\mathrm{ren}}{\delta\hat\varphi_1^*(x')\delta h_{0\mu\nu}(x)} \nn \\
=& -4k^\rho\, {\cal D}^{\mu\nu}_\rho  
  \left[ \varphi_2\,\delta(x-x') -\varphi_1 (f(x-x')-g(x-x'))\right] \nn \\
 &\ -i\varphi_1\, {\cal P}^{\mu\nu}\,(f(x-x')+g(x-x')) -2i\varphi_2\,  \frac{\partial_\mu \partial_\nu}{\Box}\,\delta(x-x') \nn \\
 &-2 i \  k^{\alpha} k^{\beta} \mathcal{T}_{\alpha \beta}^{\mu \nu} \left[c^*(x-x')+h^*(x-x') \right] \  .
\end{align} 
Notice that ${\cal D}^{\mu\nu}_\rho$ is an odd differential operator, while ${\cal P}^{\mu\nu}$ and $\mathcal{T}^{\mu \nu}_{\alpha \beta}$ are even. 
In order to obtain the expected Ward identities, one should implement the shift \eqref{eq:BFS}
\begin{equation}
\langle T_{\mu\nu}  \rangle_{QFT} = \langle T_{\mu\nu}  \rangle
+\frac 12\eta_{\mu\nu}\, (\hat \varphi_1  + \hat  \varphi_1^*)\, \langle O_\varphi \rangle =
\langle T_{\mu\nu}  \rangle +\eta_{\mu\nu}\,(\hat \varphi_1  + \hat  \varphi_1^*) \, \varphi_2 \ ,
\end{equation}
so that 
\begin{equation}
\langle T_{\mu\nu}(x)\, O_\varphi (x')\rangle_{QFT} = 2i \frac{\delta\langle T_{\mu\nu} (x) \rangle_{QFT} }{\delta \hat\varphi_1^*(x')}=
\langle T_{\mu\nu}(x)\, O_\varphi (x')\rangle +2i \varphi_2\,\eta_{\mu\nu}\, \delta(x-x')\ .
\end{equation}
Eventually we have 
\begin{align}
\langle T^{\mu\nu}(x)\, O_\varphi (x')\rangle_{QFT}
=&
 -4k^\rho \, {\cal D}^{\mu\nu}_\rho  
  \left[ \varphi_2\,\delta(x-x') -\varphi_1\, (f(x-x')-g(x-x'))\right] \nn \\
 &\ +i  {\cal P}^{\mu\nu} \left[2\varphi_2\,\delta(x-x')- \varphi_1\,(f(x-x')+g(x-x'))\right] \nn \\
 &-2 i \  k^{\alpha} k^{\beta} \mathcal{T}_{\alpha \beta}^{\mu \nu} \left[c^*(x-x')+h^*(x-x') \right] \label{TOQFT} \ .
\end{align}
The Ward identities are obtained by just taking the divergence and the trace of this expression, respectively.
From their definitions \eqref{defprojtt} and \eqref{diffops}, we know that 
$\partial_\mu {\cal D}^{\mu\nu}_\rho=\frac12 \delta^\nu_\rho$, $\partial_\mu {\cal P}^{\mu\nu}=0$, $\partial_{\mu} \mathcal{T}_{\alpha \beta}^{\mu \nu}=0$, 
$\eta_{\mu\nu}{\cal D}^{\mu\nu}_\rho=0$, $\eta_{\mu\nu} {\cal P}^{\mu\nu}=2$ and $\eta_{\mu \nu} \mathcal{T}^{\mu \nu}_{\alpha \beta}=0$.  Further using the holographic
dictionary we arrive to
\begin{align}
\langle \partial_\nu T^{\mu\nu}(x) O_\varphi (x')\rangle_{QFT} =&
 -k_\mu \langle O_\varphi \rangle\delta(x-x') \nn\\
 &-\frac{i}{2} k^{\nu} \varphi_1 \langle O_\varphi^*(x) O_\varphi(x') \rangle
 +\frac{i}{2} k^{\nu} \varphi_1 \langle O_\varphi(x) O_\varphi(x') \rangle \ ,
\label{eq:witrhol}\\
\langle T(x)O_\varphi(x')\rangle_{QFT} =&\; 2i \langle O_\varphi \rangle \delta^3(x-x') \nn\\
 &- \frac{1}{2} \varphi_1  \langle O_\varphi^*(x) O_\varphi(x') \rangle- \frac{1}{2} \varphi_1  \langle O_\varphi(x) O_\varphi(x') 
 \rangle  \ .\label{eq:widilhol}
\end{align}
These are exactly the Ward identities \eqref{2WIt} and \eqref{2ska} we anticipated from field theory, provided we recall that $O=e^{ik\cdot x}O_\varphi$, $\bar\Phi=e^{ik\cdot x}\varphi_1$ and set $d = 3$, $\Delta_O = 2$ there.

\section{The (pseudo-)Goldstone spectrum}
\label{gene} 

In Section \ref{olo} we verified that the holographic Q-lattice complies with the 
general QFT analysis of Section \ref{sec:QFT} and, specifically, it leads to the same set of Ward identities. 
We now continue to develop the generic QFT analysis of the 2-pt correlators, but we supplement it with 
some input from the holographic model. Our aim is to study general conditions that the scalar 2-pt correlators 
(\emph{i.e.} those that remain unknown before actually solving the model) have to satisfy in order to produce a consistent 
pattern of Goldstone and pseudo-Goldstone modes. One such requirement is, for instance, the absence 
of massless poles when the symmetry breaking is explicit.

In this section, we first recast the Ward identities obtained in Section~\ref{sec:QFT}
by introducing new suitable definitions.
Next, these Ward identities together with input from the holographic model help us to formulate an
Ansatz for the correlators $\langle T^{\mu \nu}(q) O_\varphi(-q) \rangle$.
An immediate goal is to uncover some relevant properties of the holographic model 
before actually solving it.
While a more ambitious objective is to try and extract some general messages from holography on the phenomenology 
of translational symmetry breaking.
In this sense, the present analysis is an initial step towards building effective field theories of
translation and dilatation 
breaking in general.

We consider the Ward identities derived in Section \ref{sec:QFT} and adopt the Q-lattice Ansatz
\begin{equation}\label{holoa}
O=e^{ik \cdot x} O_\varphi \ , \qquad \bar \Phi= e^{i k \cdot x} \varphi \ .
\end{equation}
As done in the holographic computation of the previous Section \eqref{corresca}, we introduce unknown functions to express the 2-pt correlators involving the scalar operators
\begin{subequations}
\label{eq:fgdefs}
\begin{align}
\label{f}
&\langle O_\varphi(x)\,O_\varphi^*(x') \rangle = \langle O^*_\varphi(x)\, O_\varphi(x')\rangle = 4i f(x-x')\ ,\\
\label{g}
\langle  O_\varphi(x)\, &O_\varphi(x')\rangle = 4i g(x-x') \ , \qquad 
\langle O^*_\varphi(x)\, O_\varphi^*(x')\rangle = 4i g^*(x-x') \ ,
\end{align}
\end{subequations}
where $f$ is real. 
The assumption that $f$ and $g$ are functions of the difference $x-x'$
is inspired by holography. Indeed in Section \ref{olo} we observed that
the holographic model, when written in terms of a  Q-lattice Ansatz like
\eqref{holoa} (which factors out the spatial dependence), leads to 
bulk equations in terms of the ``$\varphi$ quantities" 
that do not have any explicit space-time dependence. As a consequence, we can assume that, as a general feature, 
the correlators are effectively
insensitive to the breaking of translations. 

The 
imprint of the symmetry breaking
has been factored out by \eqref{holoa} and is encoded through the manifestly $k$-dependent 
terms in the translation Ward identity \eqref{transa} below.
This simplification  will be crucial
for the following analysis.

Adopting the redefinition \eqref{eq:BFS} for the energy-momentum tensor,
the 2-pt Ward identities of Section \ref{sec:QFT} become (we drop the $QFT$ subscript here and below)
\begin{subequations}
\label{eq:qtrWIs}
\begin{align}
\label{transa}
&\langle \partial_\mu T^{\mu \nu}(x)\, O_\varphi(x') \rangle = 
 -k^\nu\,\langle O_\varphi \rangle\, \delta(x-x') +  2k^\nu\, \varphi \left[f(x-x')-g(x-x')\right]\ ,\\
\label{dilansa}
&\langle T(x)\, O_\varphi(x') \rangle = i \Delta_O\,\langle O_\varphi \rangle\, \delta(x-x') 
 -{2i(d-\Delta_O)}\, \varphi \left[f(x-x')+g(x-x')\right]\ ,
\end{align}
\end{subequations}
and
\begin{equation}\label{rotansa}
\langle T^{\left[\mu \nu\right]}(x)\, O_\varphi(x') \rangle=0  \ .
\end{equation}

It will prove useful in the following to work in the momentum basis. We shall then
Fourier transform the Ward identitis \eqref{eq:qtrWIs} which become
\begin{subequations}
\label{eq:qtrWIsft}
\begin{align}
&q_{\mu}\langle  T^{\mu \nu}(q)\, O_\varphi(-q) \rangle = 
 ik^\nu\,\langle O_\varphi \rangle-  2ik^\nu\, \varphi \left[f(q)-g(q)\right]\ ,\\
&\langle T(q)\, O_\varphi(-q) \rangle = i \Delta_O\,\langle O_\varphi \rangle 
 -{2i(d-\Delta_O)}\, \varphi \left[f(q)+g(q)\right]\ .
\end{align}
\end{subequations}
We shall now formulate a generic ansatz for the $\langle T^{\mu \nu}(q)\, O_{\varphi}(-q) \rangle$
correlator and require it to satisfy the Ward identities above. Note first that the Ward identity \eqref{rotansa} requires the mixed 
2-pt function to be symmetric in the Lorentz indices.
Hence, the most general Ansatz that we consider  is
\begin{equation}
\label{ansatze}
\langle  T^{\mu \nu}(q)\, O_\varphi(-q) \rangle = 
A\, q^\mu q^\nu 
+ B\, \eta^{\mu\nu} 
+ C\, k^\mu k^\nu 
+ D\,(k^\mu q^\nu + k^\nu q^\mu)\ .
\end{equation}
Imposing the Ward identities \eqref{eq:qtrWIsft}, we get%
\footnote{In the case of $d=3$ and $\Delta_O=2$ one can compare the expression \eqref{corricorri} with the holographic result \eqref{TOQFT} by considering that, as one can see from \eqref{eq:fgdefs}, the functions $f$ and $g$ map to themselves while (in direct space) $C(x-x')=-2i(c^*(x-x')+h^*(x-x'))$.}
\begin{equation}
 \begin{split}\label{corricorri}
 &\langle T^{\mu\nu}(q) O_{\varphi}(-q) \rangle =
 \frac{i}{d-1}\left(\eta^{\mu\nu}-\frac{q^\mu q^\nu}{q^2}\right)
 \left[
 \Delta_O\langle O_{\varphi} \rangle  - 2  (d-\Delta_O) \varphi (f+g)+i C k^2
 \right] +Ck^{\mu}k^{\nu}\\
 &\qquad+\frac{i}{q^2} \left[
 k^\mu q^\nu + k^\nu q^\mu 
 - \frac{k\cdot q}{d-1}\left(\eta^{\mu\nu}+(d-2)\frac{q^\mu q^\nu}{q^2}\right)
 \right]
 \left[
 \langle O_{\varphi} \rangle  - 2\varphi (f-g)+ i C k\cdot q
 \right] \ .
 \end{split}
\end{equation}
The above equation is one of the main results of this work. As we now explain
it sheds light on the spectrum of (pseudo-)Goldstone modes of the model. Note that, in contrast to the relativistic invariant case, here the Ward identities do not determine completely the form of the correlator and the function $C(q)$ remains arbitrary.%
\footnote{Leaving $C(q)$ undetermined instead of another function appearing in the ansatz \eqref{ansatze} is a matter of choice. } Moreover, one salient feature is the generic presence of a $1/q^4$ term which deserves particular care and is 
analyzed in Subsection \ref{avoiding}. Besides the quartic pole, we observe that in the spontaneous case $\varphi=0$, two simple poles $1/q^2$ are present.  They represent two Goldstone modes associated 
to the phonon and the dilaton respectively.\footnote{The holographic 
dilaton has already been directly studied in \cite{Hoyos:2013gma} (see also \cite{Bajc:2013wha,Argurio:2014rja}).}
Indeed, even though a nontrivial $k^{\mu}$ 
breaks more than two generators of the conformal group, the presence of a single phonon and a 
single dilaton agrees with the Goldstone mode counting for spacetime symmetries (see \cite{Brauner:2010wm} for a review). Moreover, the dispersion relations of both the phonon and the dilaton are relativistic and
their propagation velocity (the ``speed of sound") is equal to the speed of light.
This feature can be related to the fact that we are considering a spontaneous symmetry breaking in a 
relativistic conformal field theory at zero temperature.
Finally, let us remark once again that both conformal and translation symmetries are broken by the same mechanism,
that is by the same operator acquiring a nontrivial vacuum expectation value.

At this point it is worth stressing the different nature of the relativistic Goldstone modes
we observe, arising from the spontaneous breaking of translations and dilatations in a relativistic
CFT; and those, widely studied in holography, corresponding to the spontaneous breaking
of a global symmetry (both Abelian and non-Abelian cases have been considered).
These latter modes are characteristic of superfluids and have been shown to exhibit
various types of non-relativistic dispersion relations (see for instance \cite{Amado:2013xya,Argurio:2015via}) depending on the type of superfluid. In the case of an Abelian symmetry,
their speed of sound $v_s$ is given by $v_s^2 = 1/(d-1)$ \cite{Esposito:2016ria}, and hence only at $d=2$ it is equal
to the speed of light ($c=1$). Crucially, in these setups the appearance of a time-dependent scalar profile breaks Lorentz boosts as well as the global $U(1)$ symmetry, while in the case under study a spatially modulated VEV for the scalar breaks translations and dilatations.

Now, notice that in the explicit case $\varphi \neq 0$ the theory 
should be gapped, and therefore all the massless poles, 
$1/q^2$ and $1/q^4$, have to disappear. Considering this requirement one constrains the low-$q^2$ behavior of the functions $f$, $g$ and $C$. To get more insight on this point, we notice that, since $O_\varphi$ is a complex operator, we can split it into its real and imaginary parts, $O_\varphi = O_R + i O_I$ and define (omitting the obvious dependence on $q$)
\begin{equation}
\langle O_R\, O_R \rangle = i f_R \ , \qquad
\langle O_I\, O_I \rangle = i f_I \ , \qquad
\langle O_R\, O_I \rangle = \langle O_I\, O_R \rangle =i f_o \ , 
\label{oroi}
\end{equation}
where now $f_R$, $f_I$ and $f_o$ are all real functions. They are related to $f $ and $g$ defined in \eqref{f} and \eqref{g} as
\begin{equation}
f+g=\frac{1}{2}(f_R+if_o) \ , \qquad f-g=\frac{1}{2}( f_I -i f_o)\ .
\end{equation} 
Similarly we can split the function $C$ in its real and imaginary part, $C=C_R+iC_I$. 

Requiring the theory (and specifically the correlator \eqref{corricorri}) to be gapped in the explicit $\varphi \neq 0$ case, we obtain the following conditions for 
the functions $f_o, f_R$, $f_I$, $C_R$ and $C_I$ (recall that $ik\cdot q$ is the Fourier transform of a real 
operator): 
\begin{subequations}
\label{coco}
\begin{align}\label{co}
 f_o &= - \frac{ik\cdot q}{\varphi}\ C_I(q) + {\cal O}(q^4) \ , \\ \label{co1}
 f_I  &= \frac{\langle O_\varphi \rangle}{\varphi} + \frac{ik\cdot q}{\varphi} \ C_R(q)  + {\cal O}(q^4) \ ,\\ \label{co2}
 f_R &= \frac{1}{(d-\Delta_O) \varphi} (\Delta_O \langle O_\varphi \rangle - k^2\ C_I(q) ) + {\cal O}(q^2) \ , \\ \label{co3}
 f_o  &= \frac{k^2  }{(d-\Delta_O) \varphi}\ C_R(q) + {\cal O}(q^2) \ .
\end{align}
\end{subequations}
We devote Subsection \ref{ps} to analyzing the consequences of 
these requirements.

\subsection{Pole structure}
\label{ps}
In this section, building on the results above, 
we will extract information on the low-energy behavior of the correlators.
Our analysis relies on symmetry  and consistency requirements and is not 
restricted to (the solution of) any particular model.

The contraints \eqref{coco}, supplemented by the requirement that the functions $f_o, f_R$, $f_I$, $C_R$ and $C_I$
be regular in the limit $q^2 \rightarrow 0$, imply the following low-$q^2$ behaviour for those functions:
\begin{subequations}
\label{lowqexpansion}
\begin{align}
\label{lowq}
C_I&=C_{I  (0)}+C_{I  (2)} \ q^2+\mathcal{O}(q^4) \ , \\ \label{lowq1}
C_R&=C_{R(0)}+ C_{R (2)}\ q^2+\mathcal{O}(q^4) \ , \\ \label{lowq4}
f_o&=-\frac{ik\cdot q}{\varphi}C_{I  (0)}-\frac{ik \cdot q}{\varphi}C_{I  (2)} \ q^2+\mathcal{O}(q^4) \ ,\\ \label{lowq2}
f_I&=\frac{\langle O_\varphi\rangle+ik \cdot q\ C_{R  (0)}}{\varphi}+\frac{i k \cdot q}{\varphi}C_{R  (2)} \ q^2+\mathcal{O}(q^4) \ , \\ \label{lowq3}
f_R&=\frac{\Delta_O\langle O_\varphi\rangle -k^2\ C_{I  (0)}}{ (d-\Delta_O)\varphi}+f_{R (2)} \ q^2+\mathcal{O}(q^4) \ . 
\end{align}
\end{subequations}
Note that in performing the low-$q^2$ expansion (\emph{i.e.} ``light-like" limit) the quantity $k \cdot q$ can
be safely  kept fixed since the vector $k^{\mu}$ is always aligned along one spatial direction by construction. Consistency between the conditions \eqref{co} and \eqref{co3} further imposes
\begin{equation}
k^2\ C_{R(0)} = - (d-\Delta_O)ik\cdot q\ C_{I  (0)}\ .
\end{equation}

The previous conditions constrain the system enough to determine the pole structure of the pseudo-Goldstone bosons 
associated to the explicit breaking of translations and dilatations. First of all, notice that the system \eqref{coco} 
does not constrain the leading constant value of $C_I$, $C_{I (0)}$ as well as the $q^2$ coefficients of $C_{R}$ and $C_I$, $C_{R (2)}$ and $C_{I  (2)} $. Therefore, one can make the simplifying assumption that $C_{I (0)}=C_{I(2)}=0$ so that the 
correlator $f_{o}$ vanishes to the specified order. 
From \eqref{oroi} 
we see that this assumption corresponds to considering that the physics is effectively
diagonalized in the $O_R$, $O_I$ basis.
Indeed, $f_R$, whose $q^2$ coefficient is unconstrained, becomes independent of $k$ 
and can be interpreted as the correlator containing the dilaton pole, while $f_I$, 
which still depends on $k$, accounts for the phonon contribution. 
In what follows we will analyze this simpler scenario.

Let us start considering the dilatation sector and assume the following form for the dilaton correlator $f_R$:
\begin{equation}\label{singlepoled}
f_R = \frac{\mu_D}{q^2+M_D^2}+... \ ,
\end{equation}
which features a massive simple pole in a regime where $q^2$ and $M_D^2$ are smaller than the scale
set by $\mu_D$. We also assume that there is a gap between the lowest energy mode and the rest of the spectrum. 
This is in line with known strongly coupled examples, holographic and not. Moreover, the 
form \eqref{singlepoled} 
conforms with the generic Lorentz-invariant low-energy intuition.
Next, matching with the condition \eqref{lowq3} in the low-$q^2$ limit implies:
\begin{equation}\label{GMORd}
M_D^2= \frac{(d-\Delta_O) \mu_D}{\Delta_O\langle O_\varphi\rangle}\varphi\ .
\end{equation}
Imposing that the residue $\mu_D$ neither diverge nor vanish in the $\varphi \rightarrow 0$ limit, it results that $M_D^2$ is linear in the explicit breaking parameter $\varphi$, a feature reproducing the GMOR relation for the pseudo-dilaton.

Regarding the pseudo-phonon sector, one can assume for $f_I$ a simple pole structure analogous to
\eqref{singlepoled}:
\begin{equation}\label{singlepolep}
f_I = \frac{\mu_P}{q^2+M_P^2}+... \ .
\end{equation}
The condition \eqref{lowq2} at lowest order yields:
\begin{equation}
M_P^2 = \frac{\mu_P}{\langle O_\varphi\rangle}\varphi\ ,
\end{equation}
which implies that also for the mass of the pseudo-phonon a GMOR-like relation holds. At the $q^2$ order, we have a relation between $\mu_P$ and $C_{R (2)}$:
\begin{equation}
C_{R (2)} = -\frac{\langle O_\varphi\rangle^2}{ik \cdot q \ \varphi\ \mu_P}\ .
\end{equation}
Given that $C_{R (2)}$ is otherwise unconstrained, one can easily design a $C_R$ function to match any desired $\mu_P$. The only firm requirement  on $\mu_P$ is that it should have a finite limit when $\varphi \to 0$. The limit when $k\to 0$ is also interesting, however one should pay attention to the order of limits when taking also the $\varphi \to 0$ limit. Note that since $C$, and thus $C_R$, is always multiplied by at least two powers of $k$ in the correlator \eqref{corricorri}, we can allow for inverse power dependence on $k$ in $C_{R (2)}$, also taking into account that $C_R$ can be a non-trivial function of $q^2$.

As a last comment, we point out that if one had ab-initio set $C$ to zero, which is equivalent to imposing the absence of the term 
$k^{\mu}k^{\nu}$ in the correlator $\langle T^{\mu \nu}\, O_\varphi \rangle$, the requirement \eqref{lowq2} would assume the following form:
\begin{equation}
f_I=\frac{\langle O_\varphi\rangle}{\varphi}+\mathcal{O}(q^4) \ .
\end{equation} 
The absence of a $q^2$ term in the expansion for $f_I$ would have led to an incompatibility with a single pole Ansatz as in \eqref{singlepolep}, and to 
unitarity issues if one had tried to express $f_I$ more generally in terms of a sum  of single poles.

\subsection{A closer look at the double pole}
\label{avoiding}
In order to look further into the realization
of the (pseudo-)Goldstone modes in the spectrum of our model
we will analyze more closely
the relevant low-energy correlators. Let us consider the 2-pt function in the spontaneous case

\begin{equation}
 \begin{split}\label{WIO}
 &\langle T^{\mu\nu}(q) O_{\varphi}(-q) \rangle =
 \frac{i}{d-1}\left(\eta^{\mu\nu}-\frac{q^\mu q^\nu}{q^2}\right)
 \left[
 \Delta_O\langle O_{\varphi} \rangle +i C k^2
 \right] +Ck^{\mu}k^{\nu}\\
 &\qquad\qquad+\frac{i}{q^2} \left[
 k^\mu q^\nu + k^\nu q^\mu 
 - \frac{k\cdot q}{d-1}\left(\eta^{\mu\nu}+(d-2)\frac{q^\mu q^\nu}{q^2}\right)
 \right]
 \left[
 \langle O_{\varphi} \rangle + i C k\cdot q
 \right] 
 \end{split}
\end{equation}
in greater detail.
The $k$-dependent tensorial expression in the second line can be rewritten as the sum of two terms:
\begin{equation}
\Big(k^{\mu}q^{\nu}+k^\nu q^\mu-k \cdot q\ \eta^{\mu \nu}\Big) +
\frac{d-2}{d-1 }\,k \cdot q\left(\eta^{\mu \nu}- \frac{q^{\mu} q^{\nu}}{q^2}\right)\ .
\end{equation}
Only the first term contributes to the Ward identity for translations \eqref{transa}. 
The second term is transverse, and is there just to compensate for the trace of the first. 
Notice that it is this second contribution which gives rise to the potentially problematic $1/q^4$ term in
the correlator above.

We will now see that when trying to reproduce the Ward identities from
an effective field theory approach one reaches the same conclusion as above; namely, that a term needed
to compensate for the trace of the phonon contribution introduces a 
$1/q^4$ term.

We start with dilatations.
We need a $T^{\mu\nu}$ which is linear in the dilaton,
automatically conserved, and traceless only on the dilaton EOM. The Ansatz
\begin{equation}
T^{\mu\nu} \propto (\eta^{\mu\nu}\Box - \partial^\mu \partial^\nu) \sigma
\end{equation}
is such that $\partial_\mu T^{\mu\nu}=0$ and $T\propto (d-1)\Box \sigma$. 
It is thus traceless only if $\sigma$ is a massless mode, the dilaton. 
Assuming also $ O_D\propto \sigma$, then the two-point function reads
\begin{equation}
\label{tod}
\langle T^{\mu \nu} O_D \rangle \propto \frac{i}{q^2}(q^2\eta^{\mu\nu}-q^{\mu}q^{\nu} )\ ,
\end{equation}
where we have used that $\langle \sigma \sigma\rangle = \frac{i}{q^2}$.
For translations 
we should consider a $T^{\mu\nu}$ which 
is linear in the phonon, in derivatives, and in $k^\mu$, following the spirit of \cite{Leutwyler:1996er}. It should be conserved on the phonon EOM. We try
\begin{equation}
T^{\mu\nu} \propto a\,(k^\mu \partial^\nu+ k^\nu \partial^\mu)\,\xi + 
b\, \eta^{\mu\nu} k\cdot \partial \xi\ .
\end{equation}
Conservation yields
\begin{equation}
\partial_\mu T^{\mu\nu}\propto a\, k^\nu\, \Box \xi +(a+b)\, k\cdot \partial\  \partial^\nu \xi\ .
\end{equation}
In order for the field $\xi$ to obey a single equation of motion, we must set $b=-a$ so that 
\begin{equation}
T^{\mu\nu} \propto  (k^\mu\, \partial^\nu+ k^\nu\, \partial^\mu-\eta^{\mu\nu}\, k\cdot \partial)\,\xi\ .
\end{equation}
This tensor however is not traceless:
\begin{equation}
T \propto -(d-2) k\cdot \partial \xi\ .
\end{equation}
Again, assuming $\xi$ can propagate along $k^\mu$, forces us to improve this $T^{\mu\nu}$
with a transverse part that compensates the trace:
\begin{equation}
T^{\mu\nu} \propto  \left(k^\mu \partial^\nu+ k^\nu \partial^\mu-\eta^{\mu\nu} k\cdot \partial \right)\xi + \frac{d-2}{d-1}\left(
\eta^{\mu\nu} - \frac{\partial^\mu \partial^\nu}{\Box}\right)  k\cdot \partial \xi
\ .\label{tmunuphonon}
\end{equation}
Unfortunately, this compensating part is necessarily non-local. 
We see that now assuming $O_P\propto \xi$, and given that $\langle \xi\xi\rangle= \frac{i}{q^2}$, we have
\begin{equation}
\label{top}
\langle T^{\mu \nu}\, O_P \rangle
\propto \frac{1}{q^2}\left[k^{\mu}q^{\nu}+k^\nu q^\mu-\frac{k \cdot q}{(d-1) }\left(\eta^{\mu \nu}+(d-2) \frac{q^{\mu} q^{\nu}}{q^2}\right) \right]\ .
\end{equation}
Assuming that $O_\varphi = O_D + i O_P$, we see that \eqref{tod} together with \eqref{top} have exactly the 
same structure as \eqref{WIO} if one neglects there the part proportional to the arbitrary function $C$, 
which is transverse and traceless.
This approach has shown that the double pole in $\langle T^{\mu \nu}\, O_\varphi \rangle$
can be related to the presence of a non-local term already at the level of the phonon 
energy-momentum tensor \eqref{tmunuphonon}.

There are two possible ways to address the picture we have just described. 
The first is to observe that the $1/q^4$ terms are proportional to 
$d-2$. So in a $1+1$ dimensional system they disappear. It is particularly interesting 
that in such a dimensionality
our spatially modulated scalar operator represents a lattice in all the spatial directions, and $k\cdot q$ 
is necessarily non-vanishing. 
This observation hints towards a connection between the $1/q^4$ terms in the correlator and the fact of
having spatial directions 
orthogonal to the lattice.\footnote{Since we are at zero temperature, symmetry breaking in 1+1 dimensions has in general
obstructions related to the Coleman theorem \cite{Coleman:1973ci}. However we will assume the usual argument that 
the holographic system is in the large $N$ limit, hence the theorem does not apply \cite{Witten:1978qu} (see also \cite{Anninos:2010sq,Argurio:2016xih}).}

\subsection{Avoiding the double pole?}
\label{moregeneral}

In this section we will take the Ward identities of the spontaneous case as the starting point
to try and construct an Ansatz for $\langle T^{\mu \nu} O_\varphi \rangle$ that results in this correlator
displaying a single quadratic pole.

Restricting to the spontaneous case, the Ward identities
(\ref{eq:qtrWIsft}) read 

\begin{subequations}
\label{eq:qtrWIsFT}
\begin{align}
& q^\mu \langle T_{\mu\nu}\, (q) O_\varphi(-q)\rangle = i k_\nu\langle O_\varphi\rangle\ , \\
&\langle T (q)\, O_\varphi(-q)\rangle  = i \Delta_O\,\langle O_\varphi\rangle\ .
\end{align}
\end{subequations}
As before, we adopt the generic Ansatz
\begin{equation}\label{ansfor}
\langle T_{\mu\nu} (q)\, O_\varphi(-q)\rangle = A\,q_\mu q_\nu 
+B\,\eta_{\mu\nu} +C\, k_\mu k_\nu +D\,(k_\mu q_\nu + k_\nu q_\mu)\ .
\end{equation}
Imposing the Ward identities gives
\begin{align}
& C k\cdot q + D q^2 = i \langle O_\varphi\rangle\ ,\label{trk} \\
& A q^2 + B + D k\cdot q = 0 \ ,\label{trq} \\
& A q^2 + d B + C k^2 + 2 D  k\cdot q = i \Delta_O\langle O_\varphi\rangle\ .\label{scal}
\end{align}
We now assume that $C$ has the following simple pole form
\begin{equation}
C = \frac{ i \langle O_\varphi\rangle\, \mu }{q^2 + \Pi}\ ,
\end{equation}
with $\mu$ and $\Pi$ possibly functions of $q$ and $k$. From \eqref{trk}, we see that 
\begin{equation}
D= \frac{ i \langle O_\varphi\rangle  }{q^2 } - \frac{  k\cdot q }{q^2 }C=
\frac{ i \langle O_\varphi\rangle }{q^2(q^2 + \Pi)}(q^2+\Pi - \mu\ k\cdot q)\ ,
\end{equation}
so that, if $\mu\neq 0$, $D$ has a single pole if and only if $\Pi= \mu\ k\cdot q +\mathcal{O}(q^2)$. Let us then 
set\footnote{Note that the possible $\mathcal{O}(q^2)$ term in $\Pi$ can be reabsorbed by a redefinition of $\mu$.}
\begin{equation}
C = \frac{ i \langle O_\varphi\rangle \mu }{q^2 + \mu\ k\cdot q}\ , \qquad
D =\frac{ i \langle O_\varphi\rangle  }{q^2 + \mu\ k\cdot q}\ .
\end{equation}
Then, we can see from (\ref{trq}, \ref{scal}) that $B$ is determined as 
\begin{equation}
B= 
\frac{1}{d-1}\,\frac{ i \langle O_\varphi\rangle  }{q^2 + \mu\, k\cdot q}\left[\Delta_O\, q^2 + (\Delta_O\, \mu-1)\, k\cdot q 
-\mu\ k^2\right]\ ,
\end{equation}
and it features the same pole as $C$ and $D$. Finally, $A$ is determined as
\begin{equation}
A= \frac{1}{(d-1)q^2}\,\frac{ i \langle O_\varphi\rangle  }{q^2 + \mu\ k\cdot q}\left[ -\Delta_O\, q^2 -(\Delta_O\,\mu+d-2)\, k\cdot q 
+ \mu\ k^2  \right]\ .
\end{equation}
One can cancel the additional pole in $q^2$ by having the term within the square brackets to be proportional 
to $q^2$. Given that we need to have $\mu\neq 0$, the simplest solution\footnote{More 
general solutions can contemplate a non-constant $\mu$ and also $k^2\neq 0$. It is however not immediately clear
what their physical relevance is.} is to take $k^2=0$, and
\begin{equation}
\mu=-\frac{d-2}{\Delta_O}\ .
\end{equation}
Note that for $d=2$ we have $\mu=0$, hence a usual relativistic pole at $q^2=0$, and no constraint on $k^2$.

We have thus completely characterized the Ansatz \eqref{ansfor}, which reads
\begin{equation}
\label{czero}
\begin{split}
\langle T_{\mu\nu} (q)\, O_\varphi(-q)\rangle =&\left[\frac{\Delta_O}{d-1}\,(\eta_{\mu\nu}q^2-q_\mu q_\nu  )\right.\\ 
&\left.  + k_\mu q_\nu +k_\nu q_\mu -\eta_{\mu\nu}\,k\cdot q - \frac{d-2}{\Delta_O}\, k_\mu k_\nu \right]
\frac{i\langle O_\varphi\rangle}{q^2 -\frac{d-2}{\Delta_O}\, k\cdot q}\ .
\end{split}
\end{equation}
We will now show that such a correlator can be reproduced by an effective theory where the phonon and the dilaton mix.
Let us assume the following form for a linear $T_{\mu\nu}$:
\begin{equation}
T_{\mu\nu} = \langle O_\varphi\rangle\, \frac{\Delta_O}{d-1}\,(\Box \eta_{\mu\nu}-\partial_\mu \partial_\nu)\,\sigma 
+ \langle O_\varphi\rangle\,
(k_\mu \partial_\nu + k_\nu \partial_\mu - \eta_{\mu\nu} k \cdot \partial)\,\xi+\frac{d-2}{\Delta_O}\,k_\mu k_\nu\, \sigma\ .
\end{equation}
It is conserved and traceless if $k^2=0$ and
\begin{equation}
\Box \xi+\frac{d-2}{\Delta_O}\,k \cdot \partial\sigma=0\ , \qquad 
\Box \sigma -\frac{d-2}{\Delta_O}\, k \cdot \partial\xi= 0\ .
\end{equation}
These equations are compatible with their derivation from an effective action, which necessarily 
mixes $\sigma$ and $\xi$, \emph{i.e.} the dilaton and the phonon:
\begin{equation}
\mathcal{L}_\mathrm{eff} = -\frac12 \partial \sigma \cdot \partial \sigma-\frac12 \partial \xi \cdot \partial \xi
-\frac{d-2}{\Delta_O}\, \sigma\, k \cdot \partial\xi\ .
\end{equation}
The propagators are given by
\begin{equation}
\langle \sigma \sigma \rangle = \langle \xi \xi \rangle =
\frac{i\Box}{\Box^2+\left( \frac{d-2}{\Delta_O}\right)^2 (k \cdot \partial)^2 }\ , \qquad 
\langle \sigma \xi \rangle= - \langle \xi \sigma \rangle =
\frac{ik \cdot \partial}{\Box^2+\left( \frac{d-2}{\Delta_O}\right)^2 (k \cdot \partial)^2 }\ .
\label{propag}
\end{equation}
Further assuming that $O_\varphi= \sigma + i \xi$,
we get
\begin{equation}
\begin{split}
\langle T_{\mu\nu} \, O_\varphi\rangle =&\left[ \frac{\Delta_O}{d-1}\,(\Box \eta_{\mu\nu}-\partial_\mu \partial_\nu)\right.\\ 
&\left.
+i(k_\mu \partial_\nu + k_\nu \partial_\mu 
- \eta_{\mu\nu}\, k \cdot \partial)+\frac{d-2}{\Delta_O}\,k_\mu k_\nu \right]
\frac{i\langle O_\varphi\rangle}{\Box-i\frac{d-2}{\Delta_O}\,k \cdot \partial}\ ,
\end{split}
\label{eq:tocorrk2}
\end{equation}
where we have used that $\Box^2+\left( \frac{d-2}{\Delta_O}\right)^2 (k \cdot \partial)^2
=(\Box+i\frac{d-2}{\Delta_O}k \cdot \partial)(\Box-i\frac{d-2}{\Delta_O}k \cdot \partial)$. 
In Fourier space \eqref{eq:tocorrk2} is just \eqref{czero}. We have thus been able to derive that correlator,
which features a single quadratic pole, 
from an effective action. Notice however, that we had to impose $k^2=0$ which takes us to
a scenario of unclear physical relevance. As a final observation, it is interesting to note that the mixing between the dilaton and the phonon 
does not take place for $d=2$.

Of course one could still object that requiring, as in this subsection, that there is a single 
pole in all functions is perhaps too stringent. It is indeed possible that more general Ans\"atze 
for $C$ allow for double poles which are nevertheless compatible with a structure for the 
propagator matrix similar to \eqref{propag}. It does not however seem possible to derive the generic
expression with a double pole at $q^2=0$, from a local, unitary effective field theory.

\section{Discussion and perspectives}
\label{Discu}
In the present paper we have studied the concomitant breaking of translations and conformal symmetries with the aim of investigating the general structure of phonons and their pseudo 
counterparts in the context of conformal field theory. 
This analysis constitutes a step in understanding physical systems near quantum criticality featuring the spontaneous emergence of inhomogeneous space-dependent configurations, including the spontaneous emergence of a lattice.

Inspired by the holographic Q-lattice, we have considered a scalar operator in a conformal 
field theory which spontaneously acquires an expectation value modulated along one
spatial direction. 
First, relying on QFT techniques, we analyzed the structure of the Ward identities of the model  and determined its general form. Second, turning to the holographic 
Q-lattice, we have proven that this model 
satisfies the same set of Ward identities. Holography determined the  structure of the relevant correlators which was not completely fixed by the QFT analysis. Specifically, we learned that in such model the correlators are still translational invariant, though not necessarily isotropic.

Once the structure of the correlators of the model was fixed,
we focused our attention on the features of their poles.
In particular, we have found that in the purely spontaneous case the mixed correlator 
$\langle T^{\mu \nu} O \rangle$ presents two massless single poles ($1/q^2$), 
which can be interpreted as the phonon and the dilaton respectively, and generically a massless double pole ($1/q^4$). 
The latter can be interpreted as an interplay between the phonon and the dilaton, which inevitably 
coexist in the model at hand. 
The presence of this double pole is somehow intriguing: we observed that requiring the cancellation of the double pole in the sourced case (when the
symmetry breaking is explicit) implies a non-trivial and specific coefficient of the $k^\mu k^\nu$ term in that correlator. In the spontaneous case, we have seen one instance in which the double pole disappears,
provided that $k^{\mu}$ is light-like,
$k^2=0$. Such a scenario could be interpreted as an ``infinitely boosted Q-lattice", 
although its meaning as a physical setup is not easy to ascertain, and therefore
this solution cannot be completely satisfactory.

The picture we have described already points to several possible future directions of research
that could help clarify the issue we encountered in the structure of the correlators.
The most obvious one is to solve the holographic 
model and compute the pole structure of the correlators by analyzing the quasi-normal modes.\footnote{Spatially inhomogeneous holographic systems featuring striped configurations and spontaneous translation breaking have already received much attention \cite{Donos:2011bh,Donos:2011ff,Donos:2011qt,Donos:2012gg,Donos:2012wi,Donos:2012yu,Donos:2013gda,Erdmenger:2013zaa,Donos:2014oha,Donos:2014gya,Erdmenger:2015qqa}. It would be interesting to analyze closely the relation of the present study with the existing models.}

A less obvious future direction arises from the observation that in d = 2 the double pole disappears. Even though $d = 2$ is a very peculiar case, in which at finite $N$ there is no spontaneous symmetry breaking, its study could provide useful insight about evading the double pole. It is also worthwhile analyzing models with a lattice in all the spatial directions (i.e. a Q-lattice with several scalar fields possibly coupled together) and determining if they improve the effective field theory embedding of the Q-lattice models with only one scalar operator.

Breaking of translation symmetry, explicitly or spontaneously, is also crucial in understanding transport properties in condensed matter systems.  Although there has been a concerted effort in implementing the breaking of translation symmetry in holography, be it by massive gravity models or by implementation of disorder,  our approach provides the overarching structure that governs the correlators. More importantly, contrary to the linear axion and massive gravity cases, where the breaking of translation is always explicit and might be related to the presence of a finite density of impurities in the systems \cite{Seo:2016vks}, our approach deals with spontaneous symmetry breaking of translations and provides new insight in the spontaneous emergence of phonons in the holographic context.  In this direction it is  important to further analyze   the system at finite temperature.  In fact, in the present analysis the dispersion relations of both 
the phonon and the dilaton are generically relativistic because we consider the spontaneous symmetry 
breaking of a conformal field theory at zero temperature. Studying the system at finite temperature should 
lead to a velocity of sound for the phonons different from the speed of light.\footnote{It would be interesting to investigate in the present case the possible existence of a bound on the speed of sound analogous to the one conjectured in~\cite{Cherman:2009tw}, and its possible violation along the line followed in~\cite{Hoyos:2016cob}.}

Finally, since we have argued that the double pole (\emph{i.e.} the term $1/q^4$) can be ascribed to the coexistence of the dilaton and the phonon, it would be extremely interesting to consider 
cases in which the conformal symmetry is explicitly broken before translations; 
 along this line,  it would be interesting, and also more realistic from the condensed matter point of view, to introduce a finite charge density as well.

\section*{Acknowledgments}

A.A. wants to dedicate this work to the memory of his friend \emph{Lucio Calzia}, who 
taught him the real meaning of the verb ``to fight''.

We acknowledge Javier Tarrio for collaboration at an early stage of the project. We thank Aleksandr Azatov, Matteo Baggioli, Matteo Bertolini, Francesco Bigazzi, Sananda Biswas, Simone Boi, Alessandro Braggio, Alejandra Castro, Aldo Cotrone, Paolo Creminelli,  Johanna Erdmenger, Piermarco Fonda, Victor I. Giraldo-Rivera,  Jelle Hartong, Carlos Hoyos, Nabil Iqbal, Uri Kol, Nicola Maggiore, Nicodemo Magnoli, Andrea Marzolla, Javier Mas,   Rene Meyer, Andrea Mezzalira, Ioannis Papadimitriou, Jos\'e Manuel Pen\'in, Flavio Porri, Oriol Pujolas, Alfonso V. Ramallo, Diego Redigolo, Alexandre Serantes, An\'ibal Sierra-Garcia, Riccardo Torre and Giovanni Villadoro for inspiring and cheerful conversations. The authors want to especially thank Carlos Hoyos for a crucial comment on the first version of this paper. RA is a Senior Research Associate of the Fonds de la Recherche Scientifique--F.N.R.S. (Belgium), and his research  is supported in part by IISN-Belgium (convention 4.4503.15). The work of D.M. was supported by grants FPA2014-52218-P from Ministerio de Economia y Competitividad. A.A. wants to thank the hospitality of the University of W\"{u}rzburg where part of the present work has been done. L.P.Z. is thankful to ICTP for sabbatical support and Universit\'e Libre de Bruxelles for hospitality during various stages of this work. D.A. and R.A. would like to thank SISSA and ICTP for the hospitality during completion of the work. D.A. was partially supported by the COST Action MP1210 ``The String Theory Universe''. D.A. thanks the FRont Of pro-Galician Scientists for unconditional support.

\end{document}